\documentstyle[prl,aps,preprint,epsf,floats]{revtex}
\newcommand{\beq}{\begin{equation}}
\newcommand{\eeq}{\end{equation}}
\newcommand{\bfig}[4]{\begin{figure}%
\epsfxsize #2
\hfil\epsfbox{#3}\hfill\caption{#4}\label{#1}\end{figure}}
\newcommand{\AVG}[1]{\langle{#1}\rangle}
\begin{document}
\draft
\preprint{\vbox{\hbox{UCSD/PTH 01--10}}}
\title{Global duality in heavy flavor decays in the 't~Hooft model}
\author{Benjam\'\i{}n Grinstein}
\address{Department of Physics, University of California, San Diego, 
La Jolla, CA 92093-0319}

\date{June 6, 2001}
\maketitle
\begin{abstract}
We average the decay width of a heavy meson in the t'~Hooft model over
the heavy quark mass $M$ with a smooth weight function. The
averaging has support over a few resonance spacings. We use the previously
determined heavy meson decay width which differs from the free quark width
by a $1/M$ correction. In contrast, we find that the averaged meson and quark
widths differ by a $1/M^2$ correction. We speculate on the
relevance of our results to the phenomenologically relevant case of
$3+1$ dimensional QCD.
\end{abstract}

Quark-hadron duality has been part of the lore of strong interactions
for three decades. Bloom and Gilman\cite{Bloom:1971ye,Bloom:1970xb}
(BG) discovered duality in electron-proton inelastic
scattering. There, the cross section is given in terms of two Lorentz
invariant form factors $W_1$ and $W_2$ which are functions of the
invariant mass of the virtual photon, $q^2$, and the energy transfer
to the electron, $\nu$.  Considering the form factors as functions of
the scaling variable $\omega\equiv q^2/2M\nu$, they compared the
scaling regime of large $q^2$ (and large $\nu$) with the region of
fixed, low $q^2$. They determined that, for each form factor, the low
$q^2$ curves oscillate about the scaling curve, that identifiable
nucleon resonances are responsible for these oscillations and that the
amplitude of a resonant oscillation relative to the scaling curve is
independent of $q^2$. Moreover, they introduced sum rules whereby
integrals of the form factors at low and large $q^2$ agree and noticed
that the agreement was quite good even when the integration involved
only a region that spans a few resonances.

Poggio, Quinn and Weinberg\cite{Poggio:1976af} (PQW) applied these
ideas to electron-positron annihilation. While BG compared
experimental curves among themselves, PQW compared the experimental
cross section to a scaling curve calculated in QCD. They noticed that
the weighted average of the cross section $\sigma(s)$,
\beq 
\label{eq:pqwavgdefd}
\bar \sigma(s) =  {\Delta\over\pi}
\int_0^\infty ds' {\sigma(s')\over (s'-s)^2+\Delta^2}
\eeq 
is given in terms of the vacuum polarization of the electromagnetic
current with complex argument,
\beq 
\bar \sigma(s) =  {1\over2i}\Big(\Pi(s+i\Delta)-\Pi(s-i\Delta)\Big),
\eeq 
and argued that one can safely use perturbation theory to compute this
provided $\Delta$ is large enough. This procedure was better
understood with the advent of Wilson's\cite{Wilson:1969} Operator
Product Expansion (OPE).  It is interesting to point out that the
prediction of PQW based on the two generations of quarks and leptons
known at the time did not successfully match the experimental
results. When PQW allowed for additional matter they found a best
match if they supplemented the model with a heavy lepton and a charge
$1/3$ heavy quark, anticipating the discovery of the tau-lepton and
b-quark. It is also interesting that PQW did not include in their
average of data the contribution of charm-onium resonances. When this
is done the average cross section is raised significantly at low $s$,
leaving the higher $s$ region unaffected, as shown in
Fig.~\ref{fig:PQW}.

\bfig{fig:PQW}{3.5in}{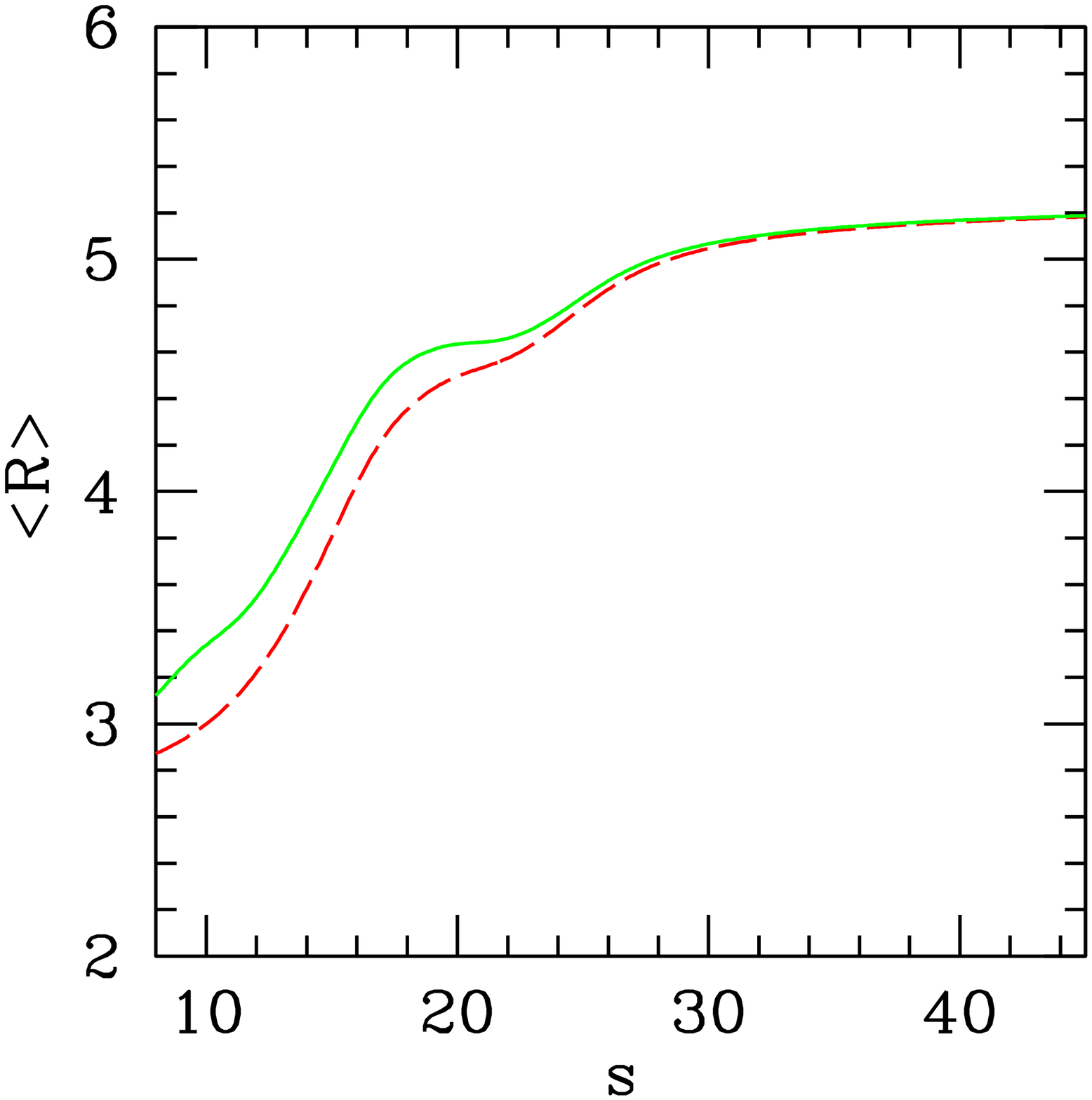}{Normalized $e^+e^-$ cross section
averaged over the squared center of mass energy $s$ as in
Eq.~(\protect\ref{eq:pqwavgdefd}) with $\Delta=3\hbox{GeV}^2$. The
dashed (red) curve is from PQW\protect\cite{Poggio:1976af}. The solid
(green) curve is ours and shows the effect of including the narrow
charm-onium resonances.}

In an attempt to understand the origin of quark-hadron duality we have
computed both the actual rate and its ``scaling limit'' from first
principles in special situations.  In Ref.~\cite{Boyd:1996ht} we
computed the semi-leptonic decay rate and spectrum for a heavy hadron
in the small velocity (SV) limit. We showed that two channels, $B\to
De\nu$ and $B\to D^*e\nu$, give the decay rate to first two orders in
an expansion in $1/m_b$ and that to that order the result is identical
to the inclusive rate obtained using a heavy quark OPE as introduced
in Ref.~\cite{Chay:1990da}. The equality holds for the double differential
decay rate if it is averaged over a large enough interval of hadronic
energies. The computation demonstrates explicitly quark-hadron duality
in semi-leptonic $B$-meson decays in the SV limit, but really sheds no
light into the mechanism for duality. In particular, it is puzzling
that duality holds even if the rate is dominated by only two 
channels.

More recently we attempted to verify duality in hadronic heavy meson
decays. In Ref.~\cite{Grinstein:1998xk} we considered the width of a
heavy meson in a soluble model that in many ways mimics the dynamics
of QCD, namely an $SU(N_c)$ gauge theory in $1+1$ dimensions in the
large $N_c$ limit. This model, first studied by
't~Hooft\cite{'tHooft:1974hx}, exhibits a rich spectrum with an
infinite tower of narrow resonances for each internal quantum number,
making the study of duality viable. We considered a `$B$-meson' with a
heavy quark $Q$ and a light (anti-)quark $q$ of masses $M_Q$ and $m$,
respectively, which decays via a weak interaction into light $\bar q
q$ mesons. To leading order in $1/N_c$ the decay rate is dominated by
two body final states: if $\pi_j$ denote the tower of $\bar q
q$-mesons, the total width is given by
$\Gamma(B)=\sum\Gamma(B\to\pi_j\pi_k)$, where the sum extends over all
pairing of mesons such that the sum of their masses does not exceed
the $B$ mass, $\mu_j+\mu_k<M_B$. The main result of that investigation
was that there is rough agreement between $\Gamma(B)$ and the decay
rate of a free heavy quark, $\Gamma(Q)$. When considered as functions
of $M_Q$ the quark rate is smooth but the meson rate exhibits sharp
peaks whenever a threshold for production of a light pair opens
up. This is due to the peculiar behavior of phase space in $1+1$
dimensions, which is inversely proportional to the momentum of the
final state mesons. Nevertheless, in between such peaks it was found
that the relation $\Gamma(B)=\Gamma(Q)(1+0.14/M_Q)$, in units of
$g^2N_c/\pi=1$, holds fairly accurately.

In this brief paper we consider the effect of local averaging on the
results of Ref.~\cite{Grinstein:1998xk}. The main result is that when averaged
locally over the heavy mass $M_Q$ the agreement between $\Gamma(B)$
and $\Gamma(Q)$ is parametrically improved. In fact, for the averaged
widths we find 
\beq
\label{eq:mainresult}
\AVG{\Gamma(B)}\approx\AVG{\Gamma(Q)}\left[1+{0.4\over M_Q^2}+{5.5\over
M_Q^3}\right] 
\eeq 
Remarkably, the correction of order $1/M_Q$ has disappeared.

The computation uses the numerical results of Ref.~\cite{Grinstein:1998xk}. The
averaging is defined by 
\beq 
\label{eq:avgdefd}
\AVG{\Gamma(M)} 
={\int_{x_{\rm min}}^{x_{\rm max}} 
dx\; x^n e^{-(x-M)^2/\sigma^2} \Gamma(x)\over
\int_{x_{\rm min}}^{x_{\rm max}} dx\; x^n e^{-(x-M)^2/\sigma^2} }.  
\eeq
The limits of integration are the lowest and highest heavy masses
available from Ref.~\cite{Grinstein:1998xk}. The width $\sigma$ was
taken to be $\sigma=1$, the scale of the strong interactions in our
units. The integer $n$ was varied between $-2$ and 1 to study the
effect of emphasizing low or high masses in the average. The result is
shown in Fig.~\ref{fig:avgwidth} where
$\log[\AVG{\Gamma(B)}-\AVG{\Gamma(Q)}]$ is plotted versus
$\log(M_Q)$. The peculiar behavior at low and large $M_Q$ (outside
$1.8<\log M_Q<2.3$) is due to the finite endpoints in the integral
defining the average as can be readily checked by averaging simple
smooth curves. The nearly straight intermediate region is therefore
what concerns us. The straight line in Fig.~\ref{fig:avgwidth} is a
linear fit by eye, which is accurate enough since there is some
variation in the results depending on the value of the power $n$ used
and what range of $\log M_Q$ is fitted. We have checked that small
variations in the linear fit do not alter our conclusions. We then
find the best fit of the exponential of this line to the function
$a+b/M_Q+c/M_Q^2$, yielding $a=-0.01$, $b=0.47$ and
$c=4.69$. The value of $a$ is consistent with zero, and the quality of
the fit is improved by dropping this degree of freedom (and gives
Eq.~(\ref{eq:mainresult})).

\bfig{fig:avgwidth}{3.8in}{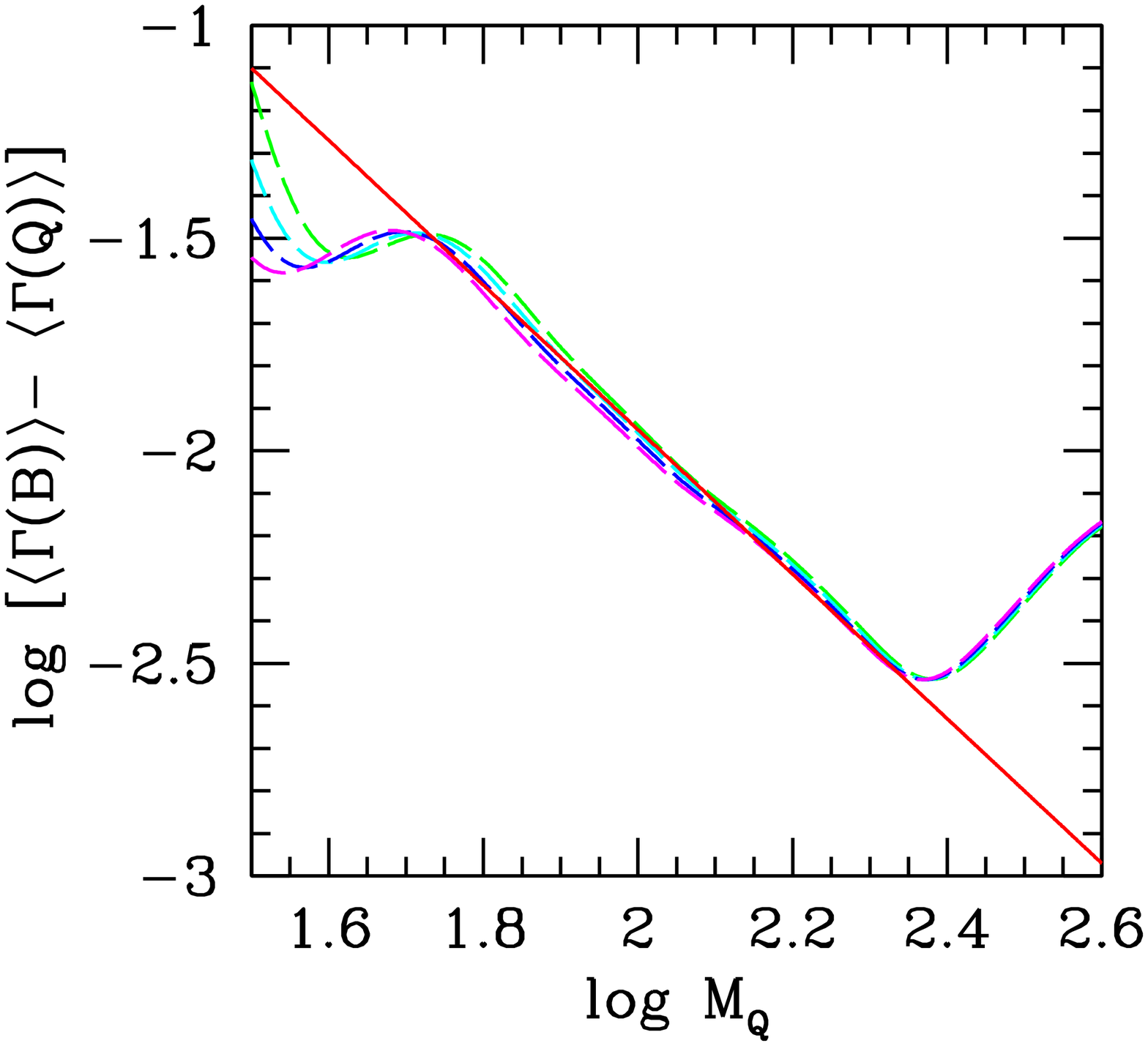}{Log-difference of
averaged meson and quark decay rates as a function of $\log
(M_Q)$ (dashed lines), and straight line fit (solid line). The different
dashed lines correspond to powers $n=-2$, $-1$, 0, and 1  in the
average of \protect Eq.~(\ref{eq:avgdefd}) ($n=-2$ is highest at small 
$M_Q$). The straight line is $-1.7\log(M_Q)+1.45$.}

Since, when plotted as functions of $M_Q$, $\Gamma(B)$ is (almost)
always above $\Gamma(Q)$ and the difference seems constant except at
the narrow peaks of $\Gamma(B)$, the question immediately arises as to
how the averaging procedure can turn the constant difference into one
that decreases as $1/M_Q$. The answer is suggested by the plot in
Fig.~\ref{fig:PQW}. Recall the average curve of PQW does not include
the effect of the narrow charm-onium resonances. Our curve includes
these narrow peaks in the averaging and when compared to the curve of
PQW it is enhanced at low $s$. The same is apparently occurring in the
averaging of $\Gamma(B)$. The area under the narrow peaks at low $M_Q$
is larger than at high $M_Q$, tilting the curve slightly.

The result of Ref.~\cite{Grinstein:1998xk} was criticized in
Ref.~\cite{Bigi:1999kc} (see also Refs.~\cite{Vainshtein:1998fg}--\cite{Bigi:1999qe}). The inconsistency between the analytic
results of Ref.~\cite{Bigi:1999kc} and the numerical results of
Ref.~\cite{Grinstein:1998xk} was blamed on numerical inaccuracies of
the latter. However, it is clear now that this is unlikely. It seems
impossible to understand what error, random or systematic,  would
conspire to change the behavior of the difference between widths from
constant for non-averaged widths to decaying as $1/M_Q$ for averaged
widths.

The main result of Ref.~\cite{Bigi:1999kc}, that there are no $1/M_Q$
corrections to the partonic width, is easy to derive. Consider the
decay $Q\to q +q'\bar q'$ where the weak interaction giving rise to
the decay is from the operator $\bar q\gamma^\mu Q\; \bar q'\gamma_\mu
q^\prime$, and take the mass of the $q'$ quarks to vanish. In this
limit the current $\bar q^\prime\gamma_\mu q'$ couples only to the
lowest meson in the tower of $\bar q'q'$ states, which is
massless. This follows from current conservation: from $\langle0|\bar
q'\gamma_\mu\gamma_5 q'|p\rangle=fp_\mu$ it follows that $fp^2=0$, so
$f\ne0$ only if $p^2=0$. It follows that the heavy meson decays into
pairs of the form $\phi\pi_n$ where $\phi$ is the massless $\bar q'
q'$ meson and $\pi_n$ are the mesons in the tower of ${\bar q}q''$
states, where $q''$ is the (light) spectator quark. The amplitude for
$B\to\phi\pi_n$ is therefore determined by the $B\to\pi_n$ form-factor
at momentum transfer $q^2=0$. The calculation of $\Gamma(B)$ is
therefore identical to the semi-leptonic width at zero $e\nu$
invariant mass. This is guaranteed to have no $1/M_Q$ corrections by
the argument of Ref.~\cite{Chay:1990da}.

Ref.~\cite{Bigi:1999kc} attempts to extend this result to the case
$m'\ne0$. This is a very delicate matter and we disagree with the
procedure presented there. In particular, the perturbative
calculations are out of control. The dimensionless expansion parameter
in the 't~Hooft model is $g^2N_c/\pi m^2$, where $m$ is the smallest
quark mass in the problem. Take for example the vertex correction to
the $\bar q Q$ current, and let $k$ be the momentum carried by the
current. If $k^2=0$ identically then, as shown in
Ref.~\cite{Bigi:1999kc}, the vertex correction vanishes. However, for
$k^2\ne0$ the leading term in the vertex correction at one loop is
proportional to $g^2N_c/\pi m^2$ (see Ref.~\cite{Callan:1976ps} and,
in particular, appendix B of Ref.~\cite{Einhorn:1976uz}). Moreover, in
the limit $k^2\to0$ the vertex correction is non-vanishing. In fact,
if all light quarks have mass $m$, the leading contribution to the
heavy quark width at one loop is (after a lengthy computation)
\beq
\Gamma(Q)^{(1-{\rm loop})}={3\over2}\left({g^2N_c\over\pi
m^2}\right)\Gamma(Q)^{(\rm tree)}.  
\eeq 
It is dangerous to study the case of non-vanishing light quark masses
as perturbations about vanishing mass.

The question immediately arises as to why is $\Gamma(Q)$ a good
approximation to $\Gamma(B)$ in the first place, since perturbation
theory clearly breaks down at small $m$. Operationally the answer to
this question is that the re-summation of the gluon ladder that gives
the quark form factor of a vector current is well approximated by the
free quark form factor. So although perturbation theory looks
hopeless, the re-summed vertex is almost like free. 

We speculate on the relevance of our results to the phenomenology
relevant case of $3+1$ dimensional QCD. The heavy meson width is not
expected to have singular peaks (as a function of $M_Q$) as in the
't~Hooft model. However, it is entirely possible that it oscillates
about the quark-dual rate. We cannot rule out, and indeed this work
suggests it is entirely possible, that the amplitude of these
oscillations decreases with $M_Q$ only as one inverse power. Once
averaged over $M_Q$ with a weight function of width of order 1~GeV the
rate may display oscillations that decrease faster with $M_Q$, say, as
$1/M_Q^2$, as apparently indicated by ``practical OPE''
arguments\cite{Bigi:1999kc}.  However, it is impractical to average over
heavy quark masses, thus leaving us with uncertainties of order
$1/M_Q$ in our predictions of hadronic widths. Some evidence for this
was presented in Ref.~\cite{Altarelli:1996gt} where it was observed that the
$b$-quark width agrees better with experimental hadronic widths if the
quark mass is replaced by the $B$ or $\Lambda_b$ masses,
respectively. In a similar vein, Ref.~\cite{Colangelo:1997ni} shows
how $1/M$ violations to local, but not global,  duality may occur in
$B$-meson correlations.

In summary, we have shown that quark-hadron duality in heavy meson
decays in the 't~Hooft model is accurate only to order $1/M_Q$, but
the accuracy is promoted to order $1/M_Q^2$ for local averages over
$M_Q$ of the widths. We speculate that the same effect occurs in four
dimensional QCD. Since averaging over $b$-quark masses is impossible,
it is safe to assume that the heavy meson and baryon widths are at
best computed with error of order $1/M_Q$.

\vskip1in
Acknowledgments. A preliminary version of these results was first
presented at Gilmanfest, Carnegie-Mellon U., May 2001. I would like to
thank the participants for their encouragement in publishing this
manuscript. This work is supported in part by the Department of Energy
under contract No.\ DOE-FG03-97ER40546.

\end{document}